\documentclass[doublecol]{epl2} 
\usepackage{amsmath,amssymb}

\newcommand{\figwidth}{0.95\columnwidth}
\newcommand{\dx}{d_{x^2-y^2}}

\newcommand{\dapical}{d_\mathrm{Cu-O}^\mathrm{\,apical}}
\newcommand{\edp}{\epsilon_d - \epsilon_p}
\newcommand{\Tcmax}{T_\text{c,max}}

\title{Scaling of the transition temperature of hole-doped cuprate superconductors with the charge-transfer energy}
\shorttitle{Scaling of cuprate transition temperatures with the charge-transfer energy} % short version since title exceeds 70 characters

\author{C. Weber\inst{1} \and C. Yee\inst{2} \and K. Haule\inst{2} \and G. Kotliar\inst{2}}
\shortauthor{C. Weber \etal}

\institute{                    
  \inst{1} T.C.M. Group - University of Cambridge, Cavendish Laboratory, J.J. Thomson Ave.,Cambridge CB3 0HE, UK \\
  \inst{2} Department of Physics \& Astronomy - Rutgers University, Piscataway, NJ 08854-8019, USA
}

\abstract{We use first-principles calculations to extract two essential
  microscopic parameters, the charge-transfer energy and the inter-cell
  oxygen-oxygen hopping, which correlate with the maximum superconducting
  transition temperature $\Tcmax$ across the cuprates.  We explore the
  superconducting state in the three-band model of the copper-oxygen planes
  using cluster Dynamical Mean-Field Theory.  We find that the variation in the
  charge-transfer energy largely accounts for the empirical trend in $\Tcmax$,
  resolving a long-standing contradiction with theoretical calculations.}

\begin{document}

\maketitle

\section{Introduction}

Despite an immense body of theoretical and experimental work, we have limited
microscopic insights of which materials-specific parameters govern the trends
in the maximum transition temperature $\Tcmax$ across the copper oxide
superconductors.  Structurally, all the cuprate families have in common CuO$_2$
planes which support superconductivity.  They are described by the chemical
formula $XS_{n-1}$(CuO$_2$)$_n$, where $n$ CuO$_2$ planes are interleaved with
$n-1$ spacer layers $S$ to form a multi-layer.  These multi-layers are then
stacked along the $c$-axis, separated by a different spacer layer $X$.
Empirically, it is known that $\Tcmax$ is strongly materials-dependent, ranging
from 40~K in La$_2$CuO$_4$ to 138~K in HgBa$_2$Ca$_2$Cu$_3$O$_8$.
Additionally, $\Tcmax$ can be tuned both as a function of doping and the number
$n$ of CuO$_2$ planes.

Studies linking the known empirical trends to microscopics have generally
established that the properties of the apical atoms (O, F or Cl, depending on
the cuprate family) are the relevant materials-dependent parameters. However,
conclusions vary regarding their effects on electronic properties, especially
in multi-layer cuprates where not all CuO$_2$ have apical atoms. Early
theoretical work by Ohta, \emph{et. al.}, found correlations between
$T_\mathrm{c}$ and the Madelung potential of the apical oxygen, arguing that
the apical potential controls the stability of the Zhang-Rice
singlets~\cite{maekawa}. They conclude that $\dapical$, the distance between
the Cu and apical O, is uncorrelated with superconductivity.  In a more recent
DFT study, Pavarini, \emph{et. al.}, argue that $\dapical$ tunes between the
single-layer cuprate families, affecting the electronic structure primarily via
the one-electron part of the Hamiltonian~\cite{scaling_Tc_andersen_prl}.
Moving the apical oxygens away from the copper oxide plane allows stronger
coupling of in-plane O $2p$ orbitals to the Cu $4s$, enhancing the strength of
longer ranged hoppings.  This effect is characterized by the increase of a
range parameter $r \sim t'/t$, describing the relative strength of the
next-nearest neighbor hopping $t'$ to nearest neighbor hopping $t$ in a
one-band model. They find that materials with larger $r$ have larger $\Tcmax$.
Many-body corrections to $t'$ were included by Yin,
\emph{et. al.}~\cite{ku_wei_apical_ref_for_dmft}.

The development of cluster Dynamical Mean-Field Theory (c-DMFT) combined with
first-principles calculations~\cite{our_review, Maier-RMP} has advanced our
qualitative and quantitative understanding of the cuprates~\cite{Gull-8site,
  Haule-SC-4site}.  A satisfactory description of these materials at
intermediate energy scales has been achieved, and the consensus is that the
cuprates lie in the regime of intermediate correlation
strength~\cite{luca_nature, our_nature_paper,
  millis_luca_analytic_continuation} near the Zaanen-Sawatzky-Allen (ZSA)
boundary~\cite{zsa_paper}.  However, all numerical studies~\cite{Kent-3band,
  tremblay_one_band_competition_antiferro, Jarrell-Hubbard-DCA} contradict the
empirical trend of $\Tcmax$ with the range parameter $r$.

\begin{figure}
  \onefigure[width=0.7\columnwidth]{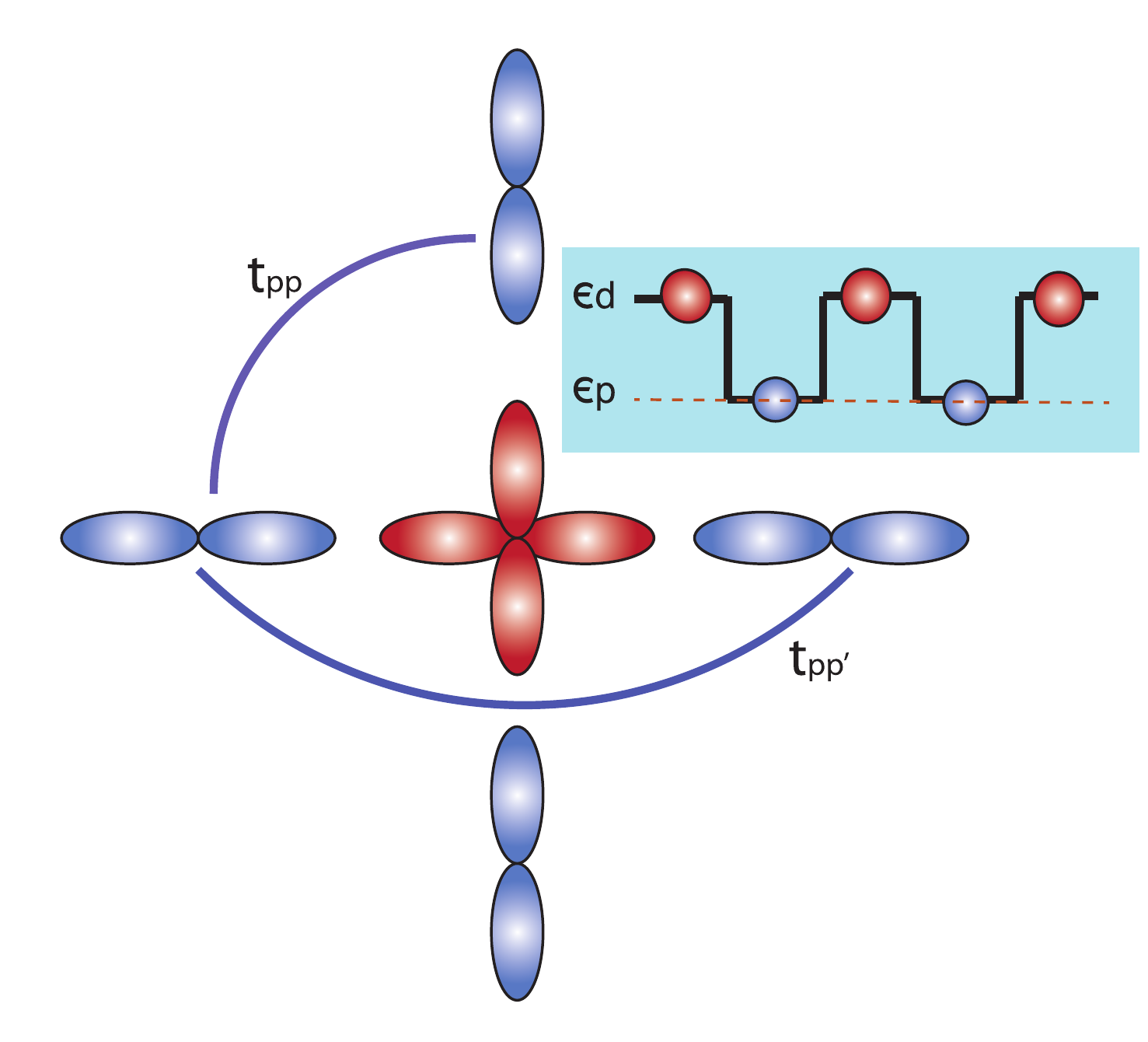}
  \caption{Parameters of the three-band $p$-$d$ model for the CuO$_2$ planes in
    the cuprate superconductors.  We show the two shortest-ranged oxygen-oxygen
    hoppings $t_{pp}$ and $t_{pp'}$, and the on-site energies $\epsilon_d$ and
    $\epsilon_p$.}
  \label{fig:3-band}
\end{figure}

In this paper, we address the origin of the variation of the experimental
$\Tcmax$ across the cuprates using recent advances in electronic structure
methods.  We carry out first-principles calculations of the hole-doped
cuprates, extract chemical parameters by downfolding to the 3-band $p$-$d$
model, and correlate them against $\Tcmax$.  Using c-DMFT, we explore the
superconducting state and identify which parameter is the key driver of
transition temperatures, resolving the conflict between numerics and the
empirical findings of Ref.~\cite{scaling_Tc_andersen_prl}.  We conclude
with suggestions for possible improvements in materials design to reach higher
critical temperatures.

\section{Trends in chemical parameters}

Effective low-energy hamiltonians containing the minimal set of bands are
important tools for understanding chemical trends.  We use the Wien2K
code~\cite{wien2k} to perform Linearized Augmented Plane Wave (LAPW)
calculations on all major copper oxide families, and then extract model
hamiltonian parameters by downfolding~\cite{lda_basis} to orbitals constructed
in the manner described in Ref.~\cite{haule-ldadmft-prb}.  In this work, we
choose to downfold to a 3-band hamiltonian describing the in-plane Cu-3$\dx$
and O-2$p$ orbitals (Fig.~\ref{fig:3-band}).  We believe four parameters
capture the essential physics: the charge-transfer energy $\edp$ between the Cu
and O atoms, the direct Cu-O hopping $t_{pd}$ and the two shortest-ranged O-O
hoppings $t_{pp}$, and $t_{pp'}$.  The extracted values are tabulated in the
Supplementary Material.

\begin{figure}
  \onefigure[width=\figwidth]{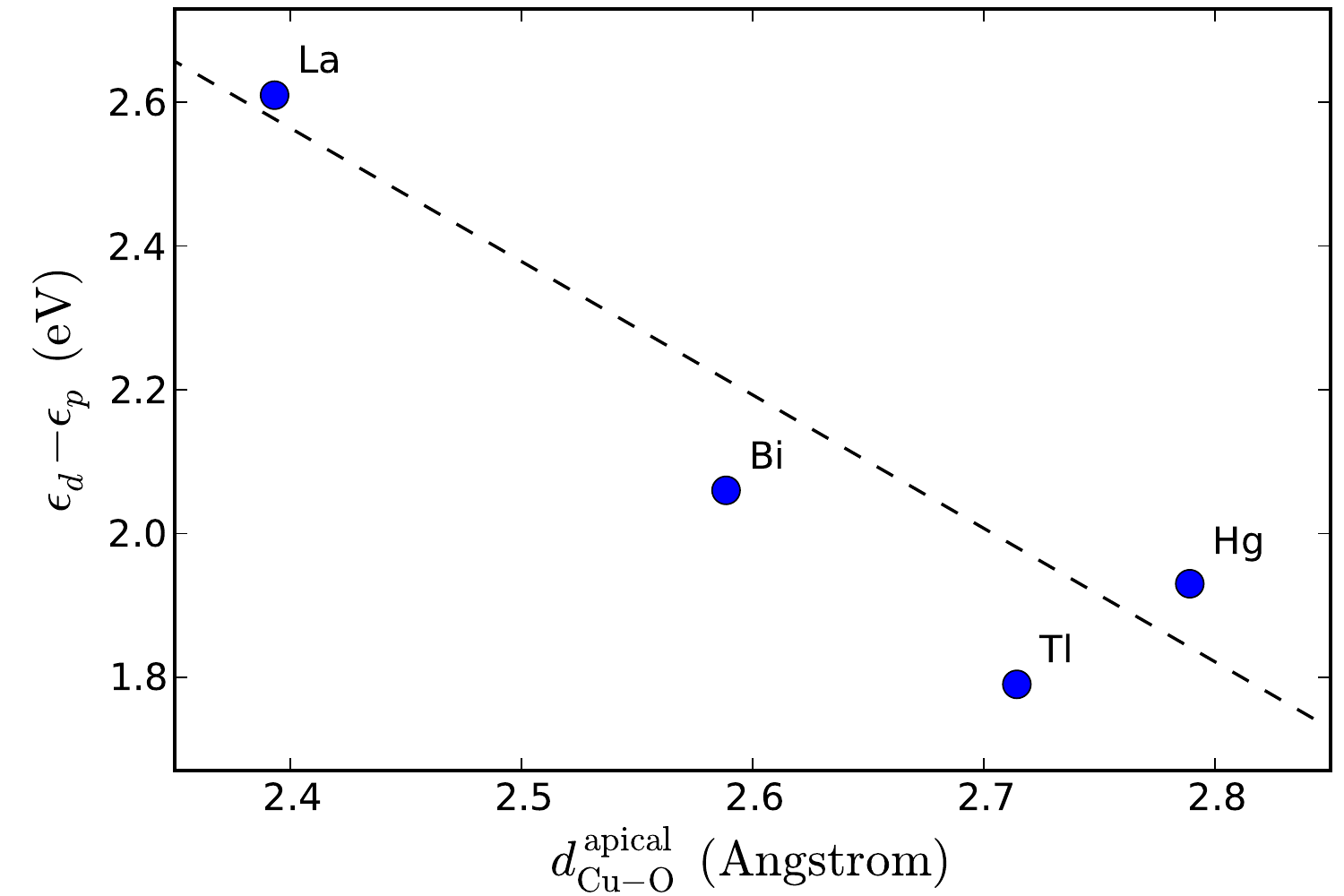}
  \caption{In single-layer cuprates, increasing the apical oxygen distance
    reduces the charge-transfer energy.}
  \label{fig:dapical}
\end{figure}

We find that only two parameters, $\edp$ and $t_{pp'}$, vary significantly
across the cuprates.  Although not crucial for our subsequent work, one would
like to have a simple structural explanation for these trends.  For the
single-layer cuprates, the variation can be directly connected to $\dapical$
(also tabulated in the Supplementary Material).  As we bring the
negatively-charged apical oxygen towards the CuO plane, the resulting
electrostatic repulsion suppresses the hopping $t_{pp'}$, since $t_{pp'}$
describes transitions of electrons past the Cu site, and provides justification
for fact that $t_{pp'}$ is smaller than $t_{pp}$~\cite{Kent-3band}.  This
mechanism for the dependence of hoppings on $\dapical$ has been pointed out in
Ref.~\cite{scaling_Tc_andersen_prl} for one-band models.  However, we show in
Fig.~\ref{fig:dapical} that the electrostatic repulsion simultaneously
increases $\edp$ by rendering it costly to place an electron on the Cu site.
These simple structural trends are less clear for multi-layer cuprates, where
additional variables such as the inter-layer distance introduce additional
complexity.

\begin{figure}
  \onefigure[width=\figwidth]{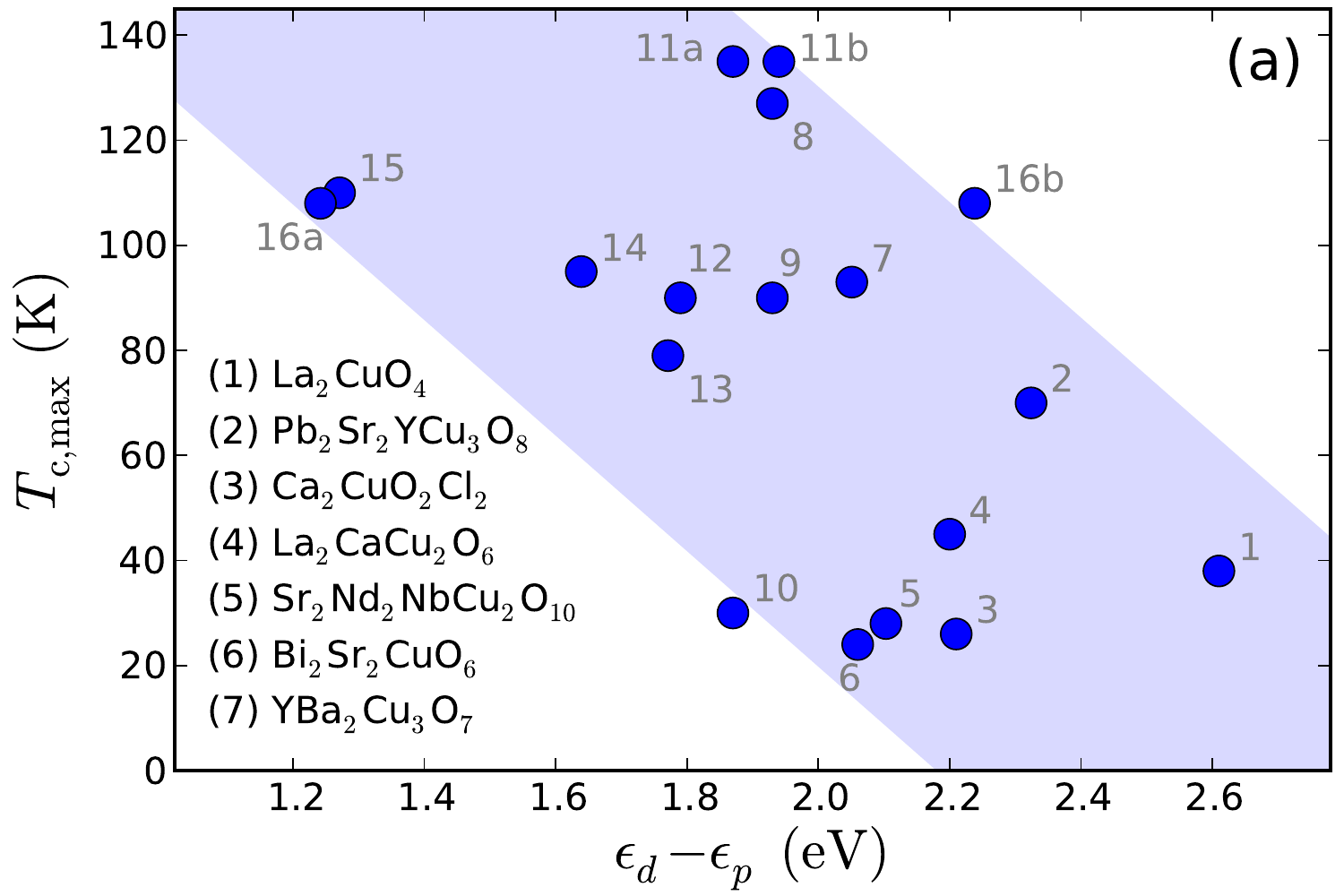}
  \onefigure[width=\figwidth]{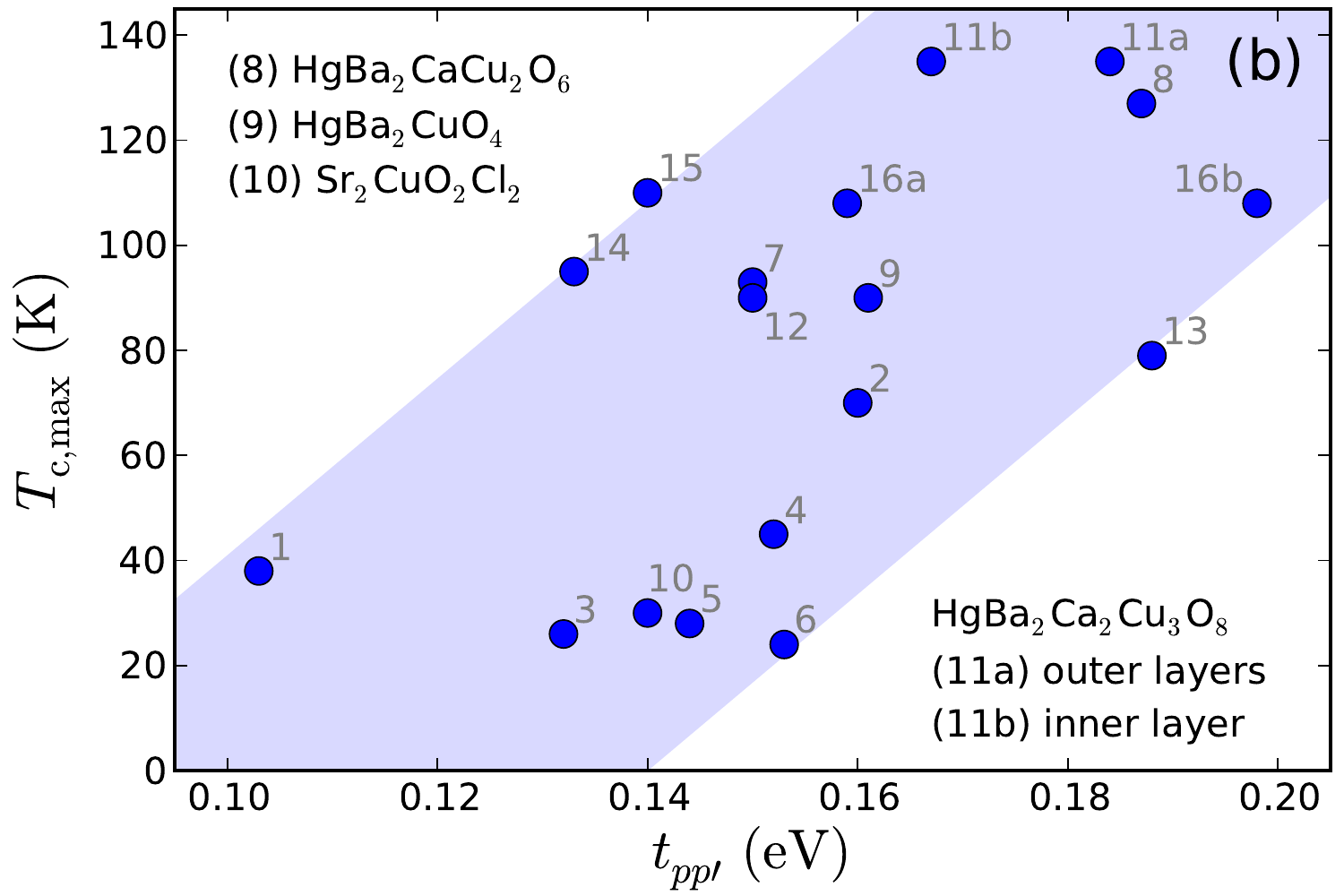}
  \onefigure[width=\figwidth]{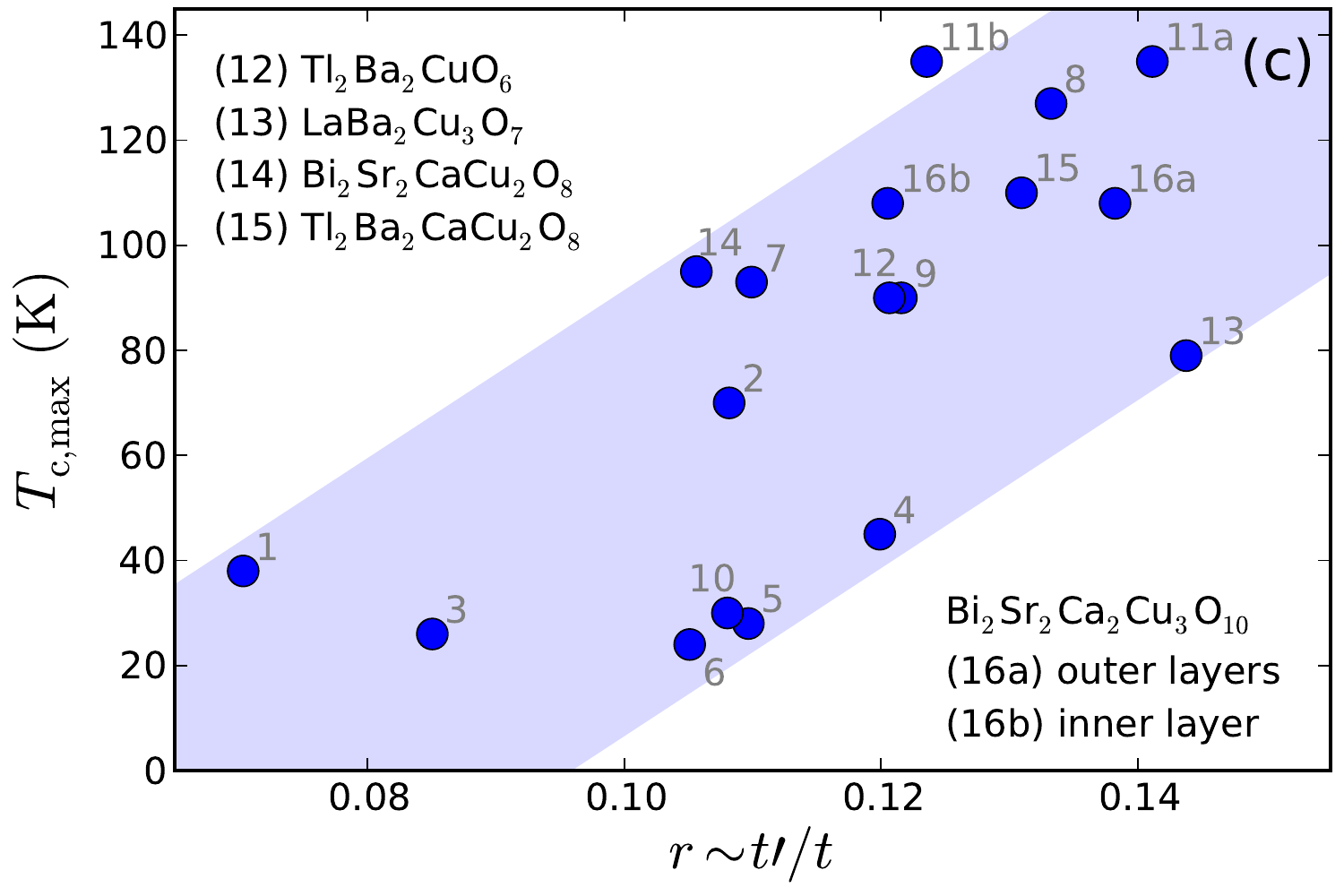}
  \caption{Correlations of $\Tcmax$ in the copper oxides with the microscopic
    parameters of the three-band model Hamiltonian with (a) the charge-transfer
    energy $\edp$ (b) the next-nearest neighbor oxygen-oxygen hopping $t_{pp'}$
    (c) the effective one-band range parameter $r \sim t'/t $.  The trend of
    the dependence of the one-band range parameter agrees with
    Ref.~\cite{scaling_Tc_andersen_prl}.}
  \label{fig:lda-Tcs}
\end{figure}

Having identified the two relevant parameters, we plot $\Tcmax$ against these
quantities in Fig.~\ref{fig:lda-Tcs}a and Fig.~\ref{fig:lda-Tcs}b to identify
possible correlations.  Beginning with La$_2$CuO$_4$ (LSCO), the limiting case
among the cuprates since it has the largest $\edp$ as well as the smallest
$t_{pp'}$, the figures show that both (i) decreasing $\edp$ and (ii) increasing
$t_{pp'}$ correlates with a enhanced $\Tcmax$.  To map our results to the
one-band Hubbard model, we integrate out the oxygen orbitals to extract the
range-parameter $r \sim t'/t$ (shown in Fig.~\ref{fig:lda-Tcs}c), and use the
fact that the effective one-band correlation strength is controlled by $\edp$
in charge-transfer materials~\cite{review_gabi}.  Our results show that both
the \emph{correlation strength} and \emph{range parameter} vary significantly
across the cuprates, in contrast with Ref.~\cite{scaling_Tc_andersen_prl} which
focused only on the latter.

\section{Correlation vs. causation}

In order to clarify how the identified microscopic parameters control $\Tcmax$,
we use c-DMFT in the cellular form~\cite{our_review, Maier-RMP} with a $2
\times 2$ cluster of impurities to solve the downfolded three-band model. The
non-local self-energy in c-DMFT captures the short-ranged correlations which
are crucial to describe $d$-wave superconductivity.  Since the fermionic minus
sign problem prevents impurity solvers based on quantum monte carlo from
accessing the low-temperature superconducting regime, we use finite-temperature
exact diagonalization (ED) at $T = 30$~K as the impurity
solver~\cite{Werner-DMFT-ED}.  In this work, we extend previous c-DMFT
calculations of the one-band model
\cite{tremblay_one_band_competition_antiferro, civelli_doping_evolution_bcs} to
the three-band model, with realistic parameters obtained from first-principles
calculations. The refinement captures the admixture of the Cu and O character
near the Fermi level via a bath representing both the Cu and O degrees of
freedom in the DMFT self-consistency condition.

The three-band hamiltonian we treat with c-DMFT is as follows:
\begin{equation}
  H = \sum_{i\alpha j\beta \sigma} t_{ij}^{\alpha\beta} c_{i\alpha\sigma}^\dagger c_{j\beta\sigma}^{}
  + \sum_{i\alpha\sigma} \epsilon_\alpha n_{i\alpha\sigma}
  + U_{dd} \sum_{i\sigma} n_{id\uparrow} n_{id\downarrow}
  \label{eq:3band-hub}
\end{equation}
where $i,j$ run over the in-plane CuO$_2$ unit cells, $\alpha,\beta$ label the
orbitals $p_x$, $p_y$ and $\dx$, and $\sigma$ is the electron spin.  The
hoppings $t_{ij}^{\alpha\beta}$ and onsite energies $\epsilon_\alpha$ are those
sketched in Fig.~\ref{fig:3-band}, except for the $d$-orbital onsite energy,
where we subtract out a doping- and material-independent double-counting
correction $E_\text{dc}$ to account for correlations included in both LDA and
DMFT.  The atomic double-counting~\cite{Anisimov-Edc}, which is very successful
for all-electron DFT+DMFT~\cite{haule-ldadmft-prb}, cannot be used because the
Wannier functions of the three-band model significantly depart from the atomic
wavefunctions.  To determine $E_\text{dc}$ for the Wannier representation, we
match the low-energy Matsubara Green's function of the three-band model to the
corresponding quantity in the \emph{ab initio} all-electron calculation (see
Supplementary Material).  A good match was attained for $E_\text{dc}=3.12$~eV
for an $\dx$ on-site Coulomb repulsion of $U_{dd} = 8$~eV.

\begin{figure}
\onefigure[width=\figwidth]{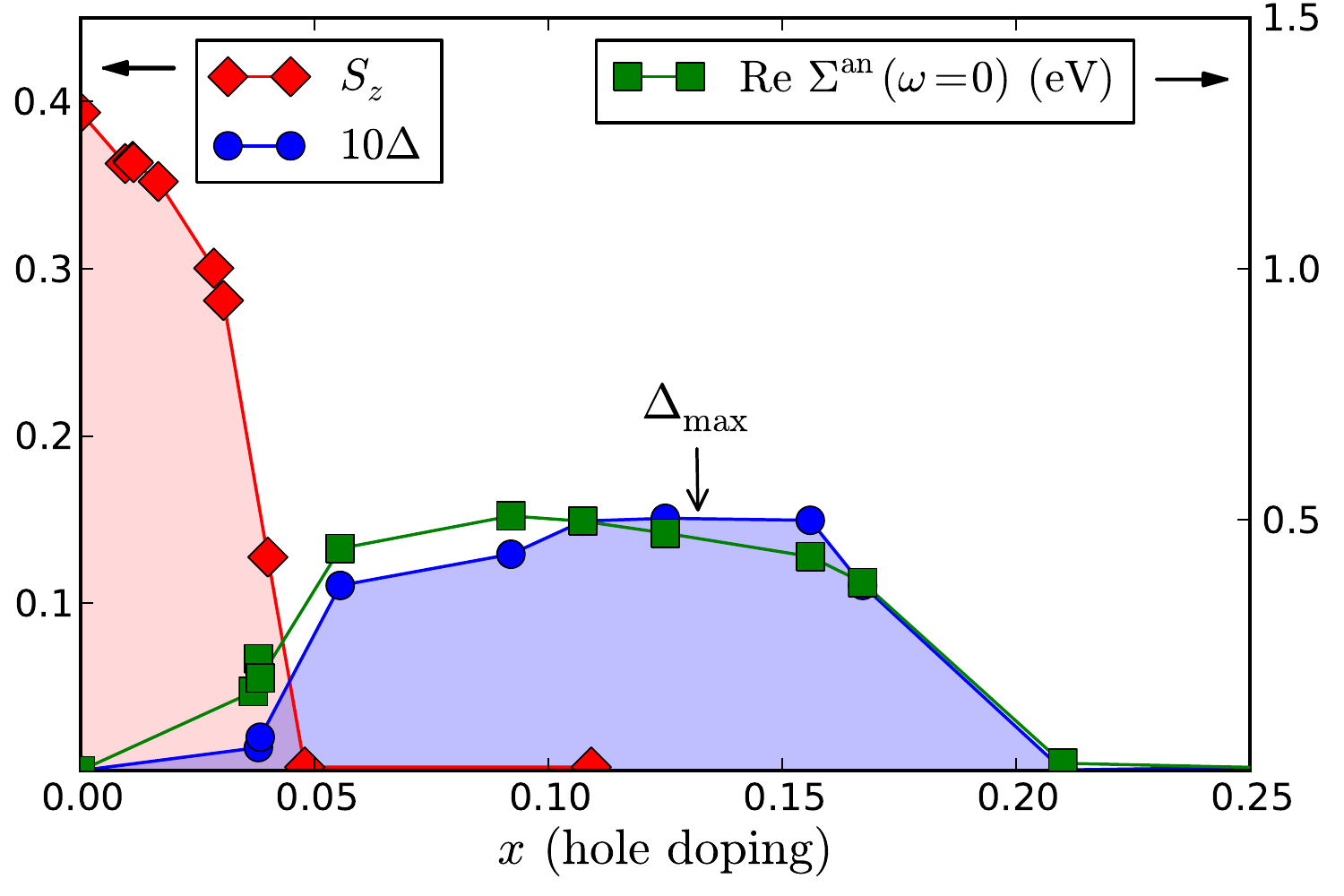}
\caption{Calculated doping dependence for LSCO of the staggered magnetization
  $S^z=\frac{1}{2}\left( n_{\uparrow} - n_{\downarrow} \right)$ and static
  $d$-wave superconducting order parameter $\Delta \sim \langle c c \rangle
  _{\tau=0}$.  We plot $10\Delta$ to fit it on the same scale as $S^z$.
  Optimal superconducting strength $\Delta_\text{max}$ is obtained for doping
  $x_{opt} \approx 0.13$.  The real part of the anomalous self-energy
  $\text{Re}\, \Sigma^{an}(\omega = 0)$ follows qualitatively the order
  parameter $\Delta$.  The calculations were performed at $T = 30$~K with
  c-DMFT and an ED impurity solver, using an 8-site discretization of the
  bath.}
\label{fig:phasediag}
\end{figure}

To test our method, we use the extracted parameters for the canonical cuprate
LSCO and explore the $T=0$ phase diagram as a function of doping. Our results,
shown in Fig.~\ref{fig:phasediag}, are qualitatively similar to experiment.
The calculations stabilize antiferromagnetism for low dopings $x<0.05$, which
gives way to a dome of $d$-wave superconductivity.  The static order parameter
$\Delta = \langle\langle c_1c_2 \rangle\rangle_{\tau=0}$, where 1 and 2 are
nearest neighbor sites on the impurity plaquette, reaches a maximum
$\Delta_\text{max}$ near $x \sim 0.13$.  We take the magnitude of
$\Delta_\text{max}$ as a proxy for the maximum superconducting temperature
$\Tcmax$.  The zero-frequency limit of the anomalous self-energy
$\Sigma^\text{an}$ is an additional indicator of superconductivity, which our
results show qualitatively follows the magnitude of the order parameter.

\begin{figure}
  \onefigure[width=\figwidth]{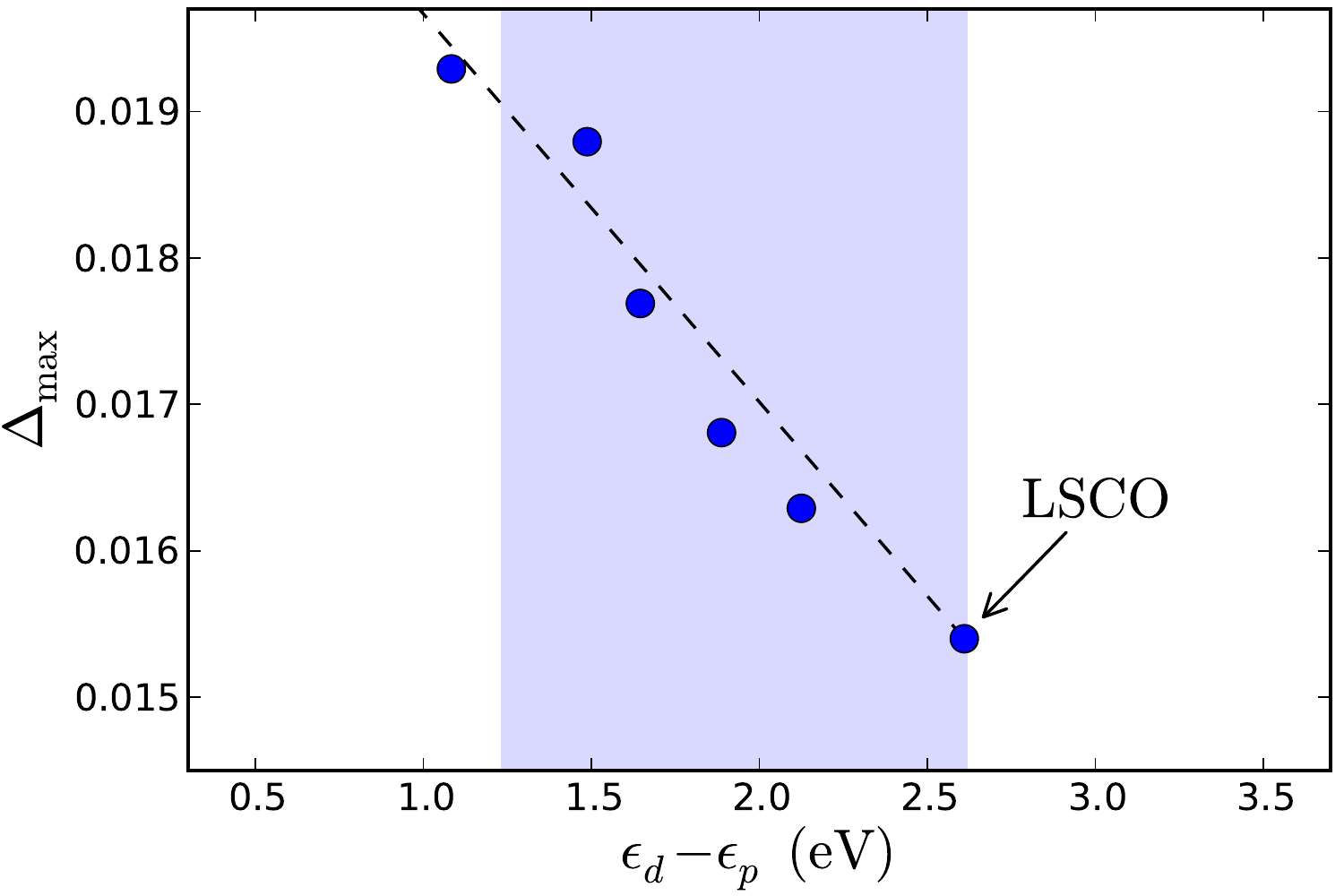}
  \onefigure[width=\figwidth]{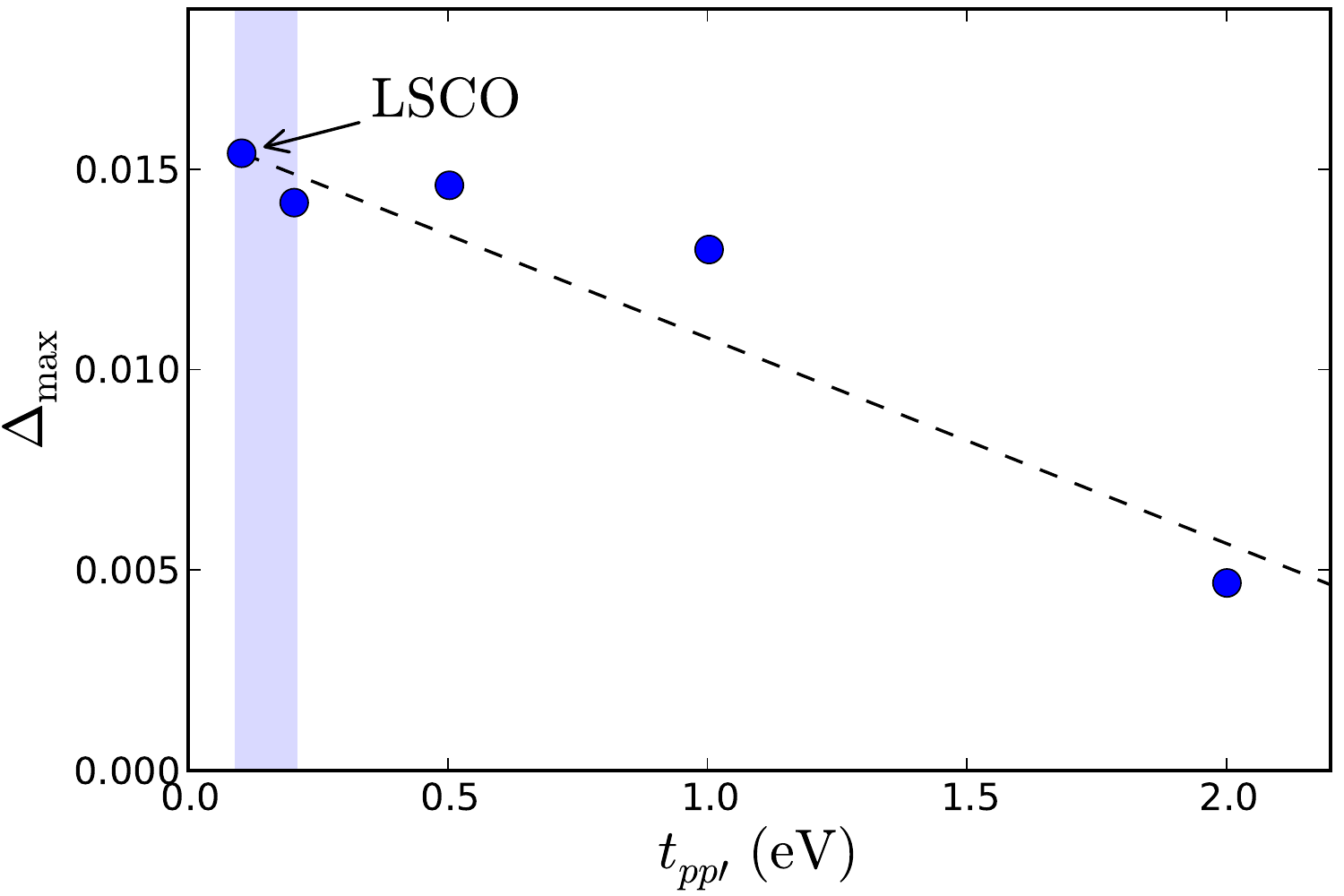}
  \caption{Optimal superconducting order parameter $\Delta_\text{max}$ of LSCO
    as we (a) decrease the charge-transfer energy $\edp$ and (b) increase
    oxygen-oxygen hopping $t_{pp'}$.  Shaded are the physical ranges spanned by
    the cuprate families.}
  \label{fig:Deltas}
\end{figure}

We argue that although \emph{two independent low-energy parameters} correlate
with the experimental $\Tcmax$, it is the charge-transfer energy that controls
the variation in $\Delta_\text{max}$, and thus $\Tcmax$, across the cuprate
families. To address this issue, we take the most correlated cuprate, LSCO, and
compute $\Delta_\text{max}$ as we either (i) decrease $\edp$ or (ii) increase
$t_{pp'}$.  Fig.~\ref{fig:Deltas}a shows that reducing the correlation strength
for fixed $t_{pp'}$ enhances the order parameter $\Delta$, in agreement with
the empirical trend in Fig.~\ref{fig:lda-Tcs}a.  However,
Fig.~\ref{fig:Deltas}b shows that increasing $t_{pp'}$ across the physical
parameter regime hardly modifies $\Delta_\text{max}$, in contrast with the
empirical trend in Fig.~\ref{fig:lda-Tcs}b.  Further increasing $t_{pp'}$ to
larger, unphysical values strongly suppresses $\Tcmax$.  Thus, our calculations
support the hypothesis that a larger hopping range $r$ suppresses $\Tcmax$, in
agreement with calculations on the
one-band~\cite{tremblay_one_band_competition_antiferro, Jarrell-Hubbard-DCA}
and three-band~\cite{Kent-3band} models.

The dependence of $\Tcmax$ on the two controlled parameters can be simply
rationalized.  For $\edp$, its large value in the strong correlation limit
suppresses charge-fluctuations, rendering the residual superexchange
interaction between the doped holes weak, resulting in low superconducting
temperatures.  As we decrease $\edp$, superconducting tendencies increase as we
pass through the intermediate correlation regime, until we reach the weak
correlation limit.  Although the ground state of the 3-band model for large
$U_{dd}$ and $\edp \sim 0$ has not been rigorously established, we expect the
large kinetic energy to suppress the effective interactions and thus
superconductivity.  Thus, we believe intermediate correlation strengths, a
regime intimately related to the charge-transfer metal-to-insulator transition,
is a crucial ingrediate for cuprate superconductivity.  Turning to $t_{pp'}$,
we find that increasing this hopping amplitude lowers the van Hove singularity
at $(0,\pi)$ away from the Fermi level.  The resulting decrease in density of
states suppresses $T_\text{c}$, an effect which simple methods
capture~\cite{kotliarsuperexchange}.  We note, however, that calculations based
on projected BCS states find the opposite trend~\cite{Randeria}, which warrants
further examination.

\section{Conclusions}

We have used electronic structure methods to identify the dependence of
$\Tcmax$ on two fundamental parameters: the charge-transfer energy $\edp$ and
inter-cell oxygen-oxygen hopping $t_{pp'}$.  We find that the position of the
apical oxygen tunes both parameters, but the strength of superconductivity,
$\Delta_\text{max}$, is mainly sensitive to $\edp$.  We expect future
refinements to explain the remaining variability in $\Tcmax$.  Our work
provides a natural interpretation of experiments where epitaxial compression in
LSCO resulted in an enhancement of $T_\text{c}$~\cite{Locquet-LSCO-pressure}.
Epitaxy increases $\dapical$ and thus reduces $\edp$.  Furthermore, our result
provides microscopic insight into the multi-layer cuprates, such as Bi-2223: in
addition to layer-dependent doping~\cite{Bi-2223-NMR}, the smaller value of the
charge-transfer energy in the outer layers may explain the enhancement of
superconductivity in the outer layers.  It has been suggested theoretically and
demonstrated experimentally~\cite{hetero-screen-Epd} that proximity to a
metallic layer reduces the charge-transfer energy.  Using this principle in
heterostructure design should result in even higher transition temperatures.

\acknowledgments

We thank A.-M. Tremblay and A. Millis for enlightening discussions.  K.H and
C.Y were supported by NSF DMR-0746395, G.K. was supported by NSF DMR-0906943,
C.W. was supported by Swiss National Foundation for Science.

\bibliographystyle{eplbib}
\bibliography{cuprates_charge_transfer}

\end{document}

\maketitle

\section{Appendix: Table of parameters}

We summarize in Table~\ref{table:lda_param} the parameters extracted via
downfolding for the three-band model and discuss the details of the downfolding
procedure.

\begin{largetable}
\begin{tabular}{cr|c|c|c|c|c|c|c|c}
  \hline
  \hline
  & Compound & $\epsilon_d-\epsilon_p$ & $t_{pd}$ & $t_{pp}$ & $t_{pp'}$ & $t'/t$ & layers & $\dapical$~(\AA) & $T_\text{c}$~(K) \\
  \hline
% Compound                                           Ed-Ep   tpd     tpp    tpp'    t'/t  Layers d(Cu-O)   Tc
  (1)   & La$_2$CuO$_4$                            & 2.61 & 1.39 & 0.640 & 0.103 &  0.070 &  1 & 2.3932 &  38 \\
  (2)   & Pb$_2$Sr$_2$YCu$_3$O$_8$                 & 2.32 & 1.30 & 0.673 & 0.160 &  0.108 &  2 & 2.3104 &  70 \\
  (3)   & Ca$_2$CuO$_2$Cl$_2$                      & 2.21 & 1.27 & 0.623 & 0.132 &  0.085 &  1 & 2.7539 &  26 \\
  (4)   & La$_2$CaCu$_2$O$_6$                      & 2.20 & 1.31 & 0.644 & 0.152 &  0.120 &  2 & 2.2402 &  45 \\
  (5)   & Sr$_2$Nd$_2$NbCu$_2$O$_{10}$             & 2.10 & 1.25 & 0.612 & 0.144 &  0.110 &  2 & 2.0450 &  28 \\
  (6)   & Bi$_2$Sr$_2$CuO$_6$                      & 2.06 & 1.36 & 0.677 & 0.153 &  0.105 &  1 & 2.5885 &  24 \\
  (7)   & YBa$_2$Cu$_3$O$_7$                       & 2.05 & 1.28 & 0.673 & 0.150 &  0.110 &  2 & 2.0936 &  93 \\
  (8)   & HgBa$_2$CaCu$_2$O$_6$                    & 1.93 & 1.28 & 0.663 & 0.187 &  0.133 &  2 & 2.8053 & 127 \\
  (9)   & HgBa$_2$CuO$_4$                          & 1.93 & 1.25 & 0.649 & 0.161 &  0.122 &  1 & 2.7891 &  90 \\
  (10)  & Sr$_2$CuO$_2$Cl$_2$                      & 1.87 & 1.15 & 0.590 & 0.140 &  0.108 &  1 & 2.8585 &  30 \\
  (11a) & HgBa$_2$Ca$_2$Cu$_3$O$_8$ (outer)        & 1.87 & 1.29 & 0.674 & 0.184 &  0.141 &  3 & 2.7477 & 135 \\
  (11b) & HgBa$_2$Ca$_2$Cu$_3$O$_8$ (inner)        & 1.94 & 1.29 & 0.656 & 0.167 &  0.124 &  3 & 2.7477 & 135 \\
  (12)  & Tl$_2$Ba$_2$CuO$_6$                      & 1.79 & 1.27 & 0.630 & 0.150 &  0.121 &  1 & 2.7143 &  90 \\
  (13)  & LaBa$_2$Cu$_3$O$_7$                      & 1.77 & 1.13 & 0.620 & 0.188 &  0.144 &  2 & 2.2278 &  79 \\
  (14)  & Bi$_2$Sr$_2$CaCu$_2$O$_8$                & 1.64 & 1.34 & 0.647 & 0.133 &  0.106 &  2 & 2.0033 &  95 \\
  (15)  & Tl$_2$Ba$_2$CaCu$_2$O$_8$                & 1.27 & 1.29 & 0.638 & 0.140 &  0.131 &  2 & 2.0601 & 110 \\
  (16a) & Bi$_2$Sr$_2$Ca$_2$Cu$_3$O$_{10}$ (outer) & 1.24 & 1.32 & 0.617 & 0.159 &  0.138 &  3 & 1.7721 & 108 \\
  (16a) & Bi$_2$Sr$_2$Ca$_2$Cu$_3$O$_{10}$ (inner) & 2.24 & 1.32 & 0.678 & 0.198 &  0.121 &  3 & 1.7721 & 108 \\
  \hline
  \hline
\end{tabular}
\caption{ Tight-binding parameters of the three-band $p$-$d$ model, containing
  the in-plane $\dx$ and $\px$ orbitals, for the hole-doped cuprates (energies
  in eV).  The table is sorted by decreasing charge-transfer energies
  $\epsilon_d-\epsilon_p$.  We have included the two nearest-neighbor
  (intra-cell) hoppings $t_{pd}$ and $t_{pp}$ as well as the inter-cell
  oxygen-oxygen hopping $t_{pp'}$.  Using the L\"owdin procedure, we have
  integrated out the oxygen bands to arrive at a one-band model, from which we
  have extracted the ratio $t'/t$ corresponding to the range parameter.  We
  also include the distance between the in-plane copper and the apical atom
  $\dapical$.  For the bilayer and trilayer compounds, we display the distance
  to the apical oxygens from the Cu atoms in the outer planes.  The last column
  displays the maximum transition temperature $\Tcmax$ for the corresponding
  optimally-doped compound.  See Ref.~\cite{scaling_Tc_andersen_prl} and
  citations therein for references to experimental work on structural
  determination and transition temperatures of the various cuprate families.}
\label{table:lda_param}
\end{largetable}

The charge-transfer energy $\epsilon_d-\epsilon_p$ is a localized, atomic-like
quantity.  Inherent in the downfolding procedure is a tradeoff between atomic
character versus faithful representation low-energy bands.  In order to
preserve as much as possible the atomic character, we implemented the first
step of the downfolding procedure described in Ref.~\cite{lda_basis}.  We
chose as initial orbitals $g_n(\mathbf{r})$ the LDA+DMFT basis constructed in
the manner described in Ref.~\cite{haule-ldadmft-prb}.  The downfolding
procedure is robust: we cross-checked our results by using
Wien2Wannier~\cite{w2w_package} and Wannier90~\cite{wannier90} to perform the
same downfolding procedure.  Our code differs slightly in the choice of radial
dependence of the trial orbitals $g_n(\mathbf{r})$.  Again, in order to
preserve the atomic character, we disabled the minimization of both spread
functionals and did not use an inner window to constrain the Fermi surface.  We
find the extracted parameters differ by less than 5\%.

In order to connect with prior work~\cite{scaling_Tc_andersen_prl}, we compute
the range parameter $r \sim t'/t$ using L\"owdin downfolding.  Beginning with
the three-band model given by $H$ equal to
\begin{equation*}
  \begin{pmatrix}
    \epsilon_d                  & 2 t_{pd} \sin\frac{k_x}{2}                    & -2 t_{pd} \sin\frac{k_y}{2}                   \\
    2 t_{pd} \sin\frac{k_x}{2}  & \epsilon_p + 2 t_{pp'} \cos{k_x}              & -4 t_{pp} \sin\frac{k_x}{2} \sin\frac{k_y}{2} \\
    -2 t_{pd} \sin\frac{k_y}{2} & -4 t_{pp} \sin\frac{k_x}{2} \sin\frac{k_y}{2} & \epsilon_p + 2 t_{pp'} \cos{k_y}              \\
  \end{pmatrix},
\end{equation*}
we integrate out the oxygen bands to arrive at the effective one-band
hamiltonian,
\begin{equation*}
  H_\text{eff}(\omega) = \epsilon_d + t_{pd} \cdot \frac{\sum_{i=0}^2 A_i(\omega) a_i(\mathbf{k})}{\sum_{i=0}^2 B_i(\omega) a_i(\mathbf{k})},
\end{equation*}
where we have defined the Fourier harmonics as
\begin{align*}
  a_0(\mathbf{k}) &= 1 \\
  a_1(\mathbf{k}) &= -2(\cos k_x + \cos k_y) \\
  a_2(\mathbf{k}) &= 4 \cos k_x \cos k_y \\
\end{align*}
and the coefficients are
\begin{align*}
  A_0 &= 4 t_{pd} (\omega - \epsilon_p + 2t_{pp})           &   B_0 &= (\omega - \epsilon_p)^2 - t_{pp}^2        \\
  A_1 &= t_{pd} (\omega - \epsilon_p + 2t_{pp'} + 4t_{pp})  &   B_1 &= t_{pp'} (\omega - \epsilon_p) - 2t_{pp}^2 \\
  A_2 &= 2 t_{pd} (t_{pp} + t_{pp'})                        &   B_2 &= t_{pp'}^2 - t_{pp}^2.
\end{align*}
Taking advantage of the fact that $B_1/B_0$ and $B_2/B_0$ are small, we expand
out the denominator and collect coefficients of the Fourier harmonics to arrive
at the range parameter
\begin{equation*}
  r \sim \frac{t'}{t} = \left. \frac{B_0 A_2 - A_0 B_2}{B_0 A_1 - A_0 B_1} \right|_{\omega = \epsilon_\text{F}}
\end{equation*}
This procedure preserves the Fermi surface and faithfully represents the
low-energy band-structure.  Since we prioritized the faithful representation of
the atomic quantities over the non-local hopping parameters, our values of $r
\sim t'/t$ are smaller than those found in Ref.~\cite{scaling_Tc_andersen_prl}.
However, the trends remain unchanged.

\section{Appendix: Numerical method}

\begin{figure}
  \onefigure[width=0.8\columnwidth]{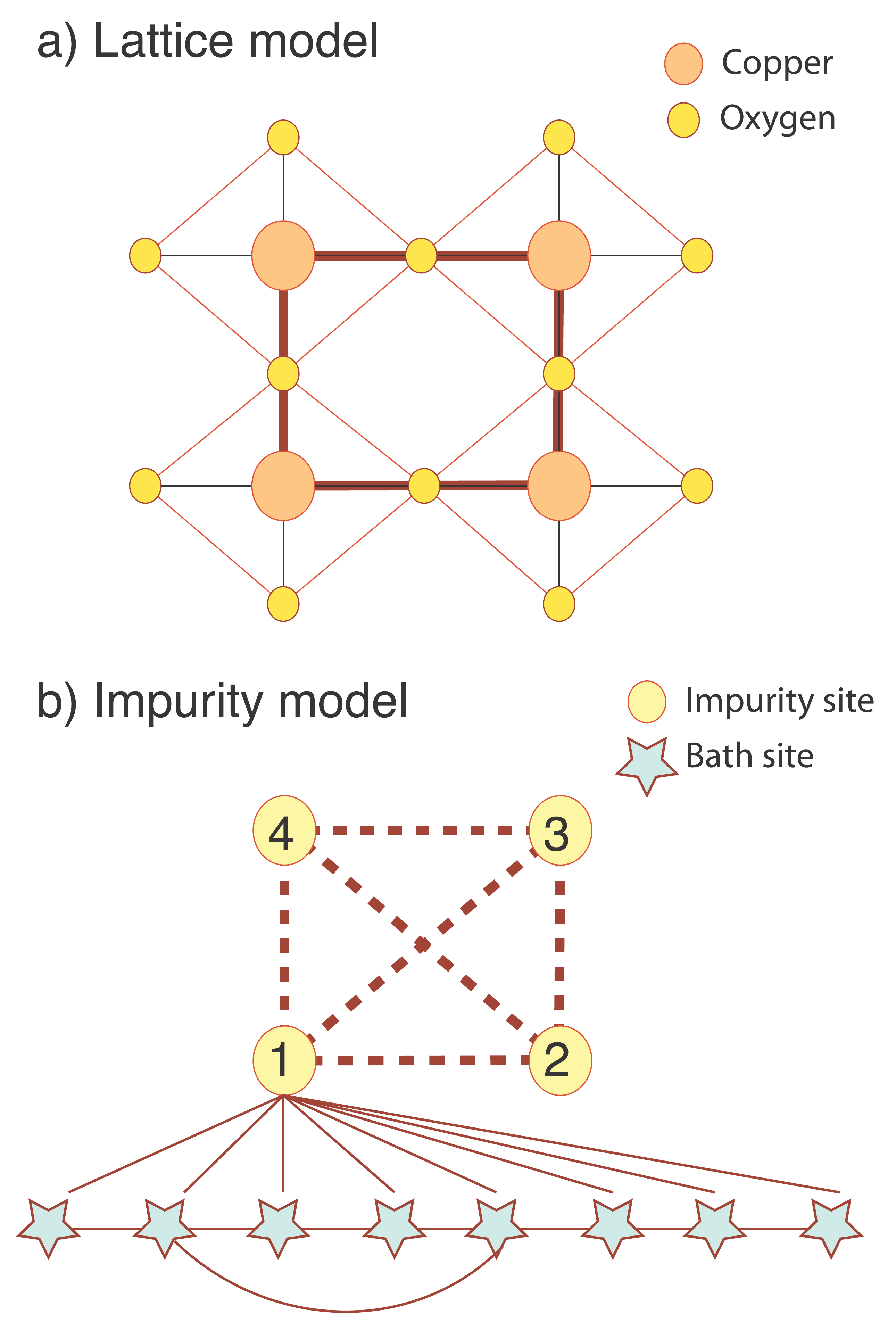}
  \caption{(a) The d-p theoretical lattice model contains the $\px$ orbitals of
    the in-plane oxygen atoms (small circles) and the $\dx$ orbital of the copper
    atoms (large circles).  The figure shows the four copper plaquette
    unit-cell used in the cellular DMFT calculations.  (b) The associated
    Anderson impurity model (AIM) contains four impurities (circles), each of
    them is independently connected to a bath discretized in eight sites
    (stars). There is not direct hybridization between the impurities, but the
    latter are connected by second order process through the bath (dashed
    lines). The sites of the bath are connected by direct and long-range
    hoppings.}
  \label{fig:model}
\end{figure}

To solve the three-band hamiltonian,
\begin{equation}
  H = \sum_{i\alpha j\beta \sigma} t_{ij}^{\alpha\beta} c_{i\alpha\sigma}^\dagger c_{j\beta\sigma}^{}
  + \sum_{i\alpha\sigma} \epsilon_\alpha n_{i\alpha\sigma}
  + U_{dd} \sum_{i\sigma} n_{id\uparrow} n_{id\downarrow}
  \label{eq:3band-hub}
\end{equation}
we use in this work the realistic set of parameters shown in
Table~\ref{table:lda_param}.  We choose a $2 \times 2$ cluster in the cellular
form~\cite{sigma_periodization_cdmft}.  Cluster DMFT improves on the single
site DMFT by adding a non-local self-energy.  In Fig.~\ref{fig:model}.{\bf a}
we show the $2 \times 2$ copper plaquette used as a unit cell through the
calculations.  The lattice Green's function of the four Copper site plaquette
is given by:
 \begin{equation}
   \label{greenfunc}
   \textbf{G}_\vk(i\omega_n) =  \left( i\omega_n + \mu - \mathbf{H}_\vk -\mathbf{\Sigma}(i\omega_n) \right)^{-1},
 \end{equation}
where $\mathbf{H}_\vk$ is the Fourier transform of the uncorrelated part of the
Hamiltonian defined in Eq.~\ref{eq:3band-hub}.  $\mathbf{\Sigma}$ is the
cluster self-energy matrix being nonzero only for the matrix elements
connecting the $\dx$ orbitals.

The self energy matrix in Eq.~\ref{greenfunc} is obtained by solving an $2
\times 2$ impurity Anderson model (shown in Fig.~\ref{fig:model}b) subject to
the DMFT self-consistency condition:
\begin{equation}
  \label{self}
  i\omega-E_{imp}-\mathbf\Sigma(i\omega)-\mathbf\Delta(i\omega)  = \hat P \left(\sum\limits_{\vk}  {G_\vk (i\omega)} \right)^{-1},
\end{equation}
where the sum runs over the reduced Brillouin Zone (BZ), and $\hat P$ is
projecting the averaged green function onto the impurity cluster subspace.

In this work we use the exact diagonalization impurity solver
algorithm~\cite{Werner-DMFT-ED}.  To solve the cluster impurity problem, we
express it in the form of a Hamiltonian $H_{imp}$ with a discrete number of
bath orbitals coupled to the cluster and use the Lanczos algorithm to converge
the ground state of the Hamiltonian and the lower states of the spectrum.  The
ED method in conjunction with c-DMFT has been widely used for the one-band
model~\cite{civelli_doping_evolution_bcs,
  tremblay_one_band_competition_antiferro, leibsch_ed_lanczos_dmft}.

The Anderson Impurity Model (AIM) is defined by :
\begin{equation}\label{aim}
  \begin{split}
    H_\text{imp} &=
    \sum\limits_{mn\sigma} \left( \epsilon^{n}_{mn\sigma}a^\dagger_{m\sigma}a_{n\sigma}
    + \epsilon^{a}_{mn\sigma}a^\dagger_{m\sigma}a^\dagger_{n-\sigma} + \text{h.c.}\right) \\
    &+ \sum\limits_{mi\sigma}{V_{mi\sigma}(a^\dagger_{m\sigma}c_{i\sigma}+h.c.)} \\
    &+ \mu \sum\limits_{i\sigma}{ c^\dagger_{i\sigma}c_{i\sigma}} +
    \sum\limits_{i\sigma}{U \hat n_{i\uparrow} \hat n_{i\downarrow}}.
  \end{split}
\end{equation}
The fermionic operators $a^\dagger_{mn}$ ($a_{mn}$) creates (destroys) a
particle in the bath, and the fermionic operators $c^\dagger_i$ ($c_i$)
creates (destroys) a particle in the cluster of impurities.  The indices $m,n$
are running over the bath sites, and the index $i$ is running over the impurity
sites, the sites of the bath are connected by long-range hopping matrix
elements through the particle-hole (particle-particle) channel $\epsilon^{n}$
($\epsilon^{a}$), the non-correlated sites of the bath are also connected to
the correlated impurities by the matrix elements $V_{mi}$, the onsite repulsion
at the impurity sites is $U$ (equal to the Coulomb repulsion of the Copper site
$U_d$), and $\mu$ is the chemical potential. We define $\epsilon$ as the
extended matrix which contains the normal $\epsilon^n$ and anomalous
$\epsilon^a$ blocks in the Nambu basis:
\begin{equation}
  \epsilon = \begin{pmatrix}
    \epsilon^n & \epsilon^a \\
    (\epsilon^a)^T & -(\epsilon^n)^T
  \end{pmatrix}
\end{equation}
The Weiss field $\mathcal{G}(i\omega_n)=i\omega_n-\Delta(i\omega_n)-E_{imp}$ is
constructed from the parameters of the AIM:
\begin{equation}
\label{weiss}
\mathbf\Delta(i\omega_n)=\mathbf{V}^\dagger \left(  i\omega_n -  \mathbf{\epsilon} \right)^{-1} \mathbf{V}
\end{equation} 
The parameters of Eq.~\ref{aim} are determined by imposing the self-consistency
condition in Eq.~\ref{self} using a conjugate gradient minimization algorithm:
\begin{equation}
  \label{distance}
  d = \sum\limits_{\omega<\omega_0,\alpha\beta}
  \left| \Delta_{\alpha\beta}^{ED}(i\omega_n) - \Delta_{\alpha\beta}(i\omega_n) \right|^2,
\end{equation}
where $\alpha\beta$ run over the matrix elements, $\omega_0 = 20$~eV is a hard
cutoff on the summation and $\Delta^{ED}$ is a function (Eq.~\ref{weiss}) of
the Hamiltonian parameters. The fitting procedure is not exact due to the
discretization of the bath and is an additional approximation to the c-DMFT
scheme.  In this work we considered a bath discretised with 8 energy levels.
Finally, once the Hamiltonian parameters are obtained by the fitting procedure,
we obtain the low energy spectrum by the Lanczos procedure.  We impose an
energy cutoff $E_{max}$ such that the Boltzman weight
$e^{-\beta(E_{max}-E_0)}<0.001$, where $E_0$ is the ground state energy.  We
discard all the eigenstates which have an energy larger than the cutoff
$E_i-E_0>E_{max}$. Once the eigenstates are obtained we compute the Boltzman
weighted average to get the dynamical and static observables.

In this work we consider two different instabilities: i) the superconducting
phase, and ii) the long-range magnetic ordered phase.  The former is computed
in the Nambu basis, and the Hilbert space is block diagonalized by the spin
$S^z$ quantum numbers, and the latter is obtained in the tensor product of the
up and down spins.  Since the number of particles is not a good quantum number
in the superconducting phase, we work at fixed chemical potential.  For the
magnetic phase we found a better convergence when working at fixed density,
with a free chemical potential.

For the determination of the phase diagram we used physical observables readily
available from the $2 \times 2$ cluster of impurities, such as the staggered
magnetization $S^z=\frac{1}{2}\left( n_{\uparrow} - n_{\downarrow} \right) $,
the superconducting order parameter $\Delta= \langle \langle c_1 c_2 \rangle
\rangle _{(\tau=0)}$ (where $1$ and $2$ are nearest neighbor links of the
impurity plaquette) and the anomalous self-energy at zero frequency
$\Sigma^\text{an} \equiv \Sigma^\text{an}_{12}(\omega = 0)$.  We emphasize that
the computed order parameters do not rely on any additional procedure, such as
the $\Sigma$-periodization \cite{sigma_periodization_cdmft}, which interpolates
and extrapolates the discrete cluster quantities to the continuum in k-space.

\begin{figure}
  \onefigure[width=0.95\columnwidth]{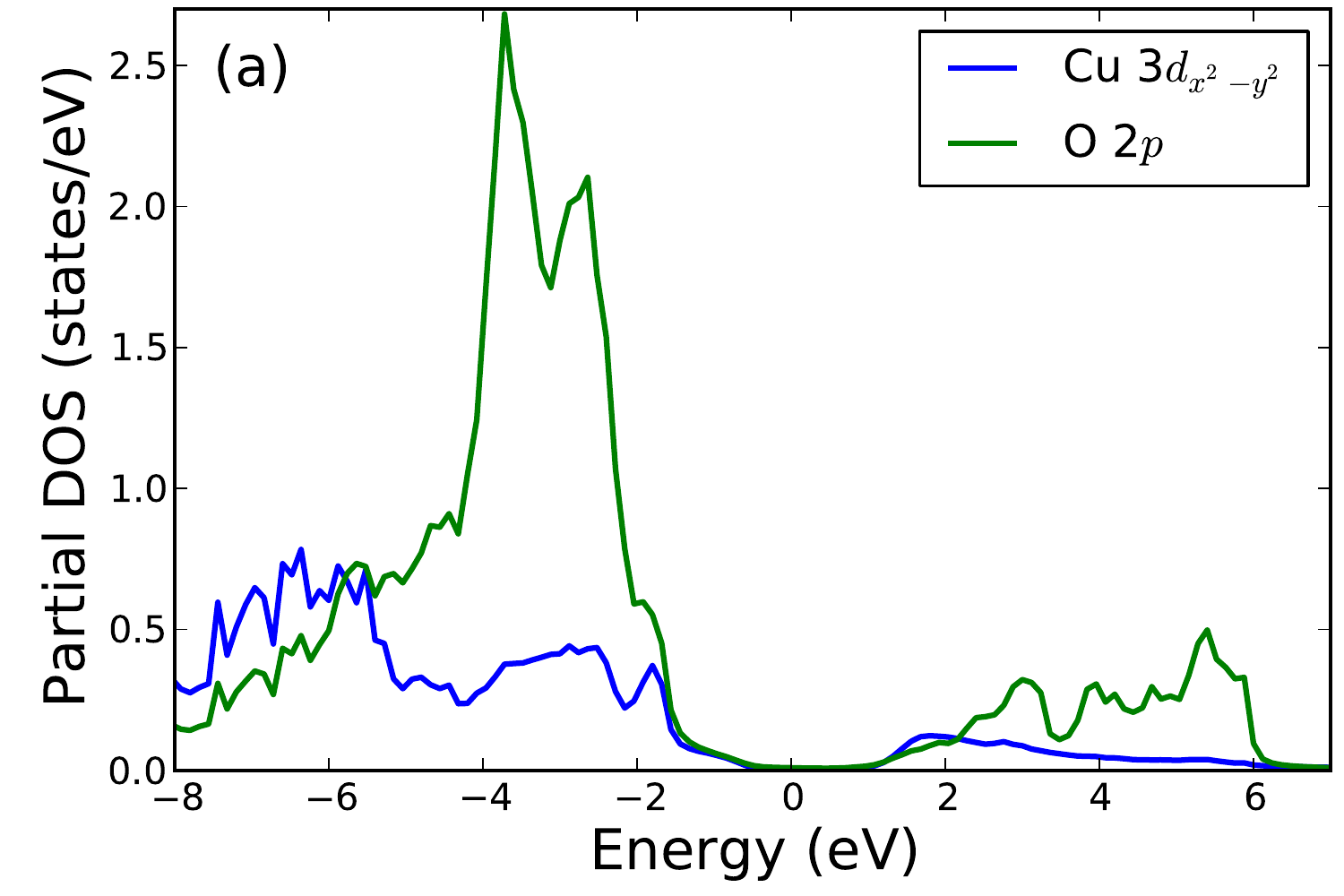}
  \onefigure[width=0.95\columnwidth]{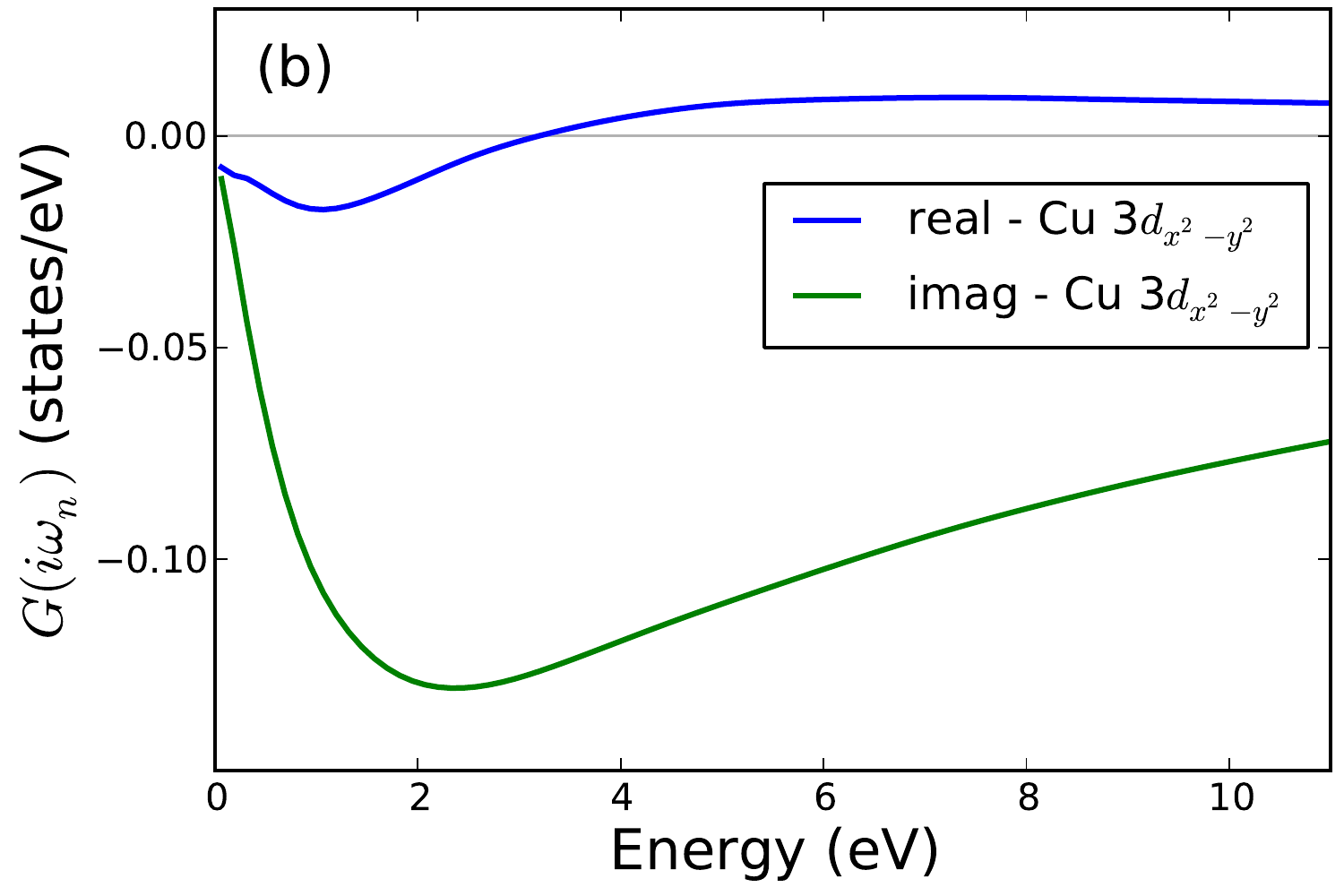}
  \caption{(a) Density of states of LSCO from the all-electron DFT+DMFT
    calculation.  (b) All-electron Matsubara Green's function for the Cu
    3$d_{x^2-y^2}$ orbital used as reference to fix double-counting in
    three-band $p$-$d$ model.  The temperature is $\beta = 50$~eV${}^{-1}$.}
  \label{fig:LSCO}
\end{figure}

Finally, we discuss the computation of the double-counting correction
$E_\text{dc}$.  Since the Wannier functions of the three-band $p$-$d$ model is
not atomic-like, we cannot use the atomic double-counting proposed in
Ref.~\cite{Anisimov-Edc}.  Rather, we first perform the \emph{ab initio}
all-electron calculation using the atomic double-counting,
\begin{equation}
  E_\text{dc} = U \left(n_{d0} - \frac{1}{2} \right) - J\left( \frac{n_{d0}}{2} - \frac{1}{2} \right),
\end{equation}
with $n_{d0} = 9$, the natural value derived from chemical valence counting.
We use $U = 10$~eV and $J = 0.7$~eV.  The atomic form of the double-counting is
appropriate here because treating the full energy window causes the orbitals to
be very atomic-like.  In Fig.~\ref{fig:LSCO}a, we plot the density of states
from the all-electron calculation, which exhibits the charge-transfer gap of
the correct magnitude (slightly less than 2~eV).  Then, we select the
double-counting in the $p$-$d$ model so the Matsubara Green's function matches
the corresponding quantity in the all-electron calculation
(Fig.~\ref{fig:LSCO}b).  We find that $E_\text{dc} = 3.12$~eV gives a good
match, and use this value for all subsequent model calculations.  Finally, we
use a reduced onsite-repulsion $U_\text{3-band} \approx U - 2J$ for our 3-band
calculations to capture the effect of the Hund's coupling present in the
all-electron calculation.

\bibliographystyle{eplbib}
\bibliography{cuprates_charge_transfer}